%% file: acl_latex.tex
\documentclass[11pt]{article}
\usepackage{booktabs}
\usepackage[table]{xcolor}
\usepackage[preprint]{acl}
\usepackage{times}
\usepackage{latexsym}
\usepackage{amsmath}
\usepackage[T1]{fontenc}

\usepackage[utf8]{inputenc}
\usepackage{enumitem}
\usepackage{microtype}

\usepackage{inconsolata}

\usepackage{fvextra}
\usepackage{tcolorbox}
\tcbuselibrary{breakable,skins}

\usepackage[table]{xcolor}  
\usepackage{soul}       
\usepackage{booktabs}
\usepackage{array}
\usepackage{caption}
\usepackage{soul}
\usepackage{longtable} 
\definecolor{highlight}{RGB}{255,255,180}
\definecolor{darkhighlight}{RGB}{255,230,130}
\definecolor{redacted}{RGB}{0,0,0}
\definecolor{tablegray}{RGB}{245,245,245}
\newcommand{\hltext}[1]{{\sethlcolor{highlight}\hl{#1}}}
\newcommand{\dhltext}[1]{{\sethlcolor{redacted}\hl{#1}}}

\newtcolorbox[auto counter, number within=section]{NewBox}[2]{%
  breakable, width=\textwidth,
  colback=red!1, colframe=red!55!black,
  colbacktitle=red!6, coltitle=black,
  fonttitle=\bfseries,
  boxrule=1.0pt,
  leftupper=0.5em, rightupper=0.5em,
  title={#1},
  label={#2},
}

\newtcolorbox{NewBoxFloat}[2]{%
  width=\linewidth,
  colback=red!1, colframe=red!55!black,
  colbacktitle=red!6, coltitle=black,
  fonttitle=\bfseries,
  boxrule=1.0pt,
  leftupper=0.5em, rightupper=0.5em,
  title={#1},
  label={#2},
}

\usepackage{graphicx}

\title{Agent Skills: A Data-Driven Analysis of Claude Skills for Extending Large Language Model Functionality}

\author{
 \textbf{George Ling\textsuperscript{1}},
  \textbf{Shanshan Zhong\textsuperscript{2}},
  \textbf{Richard Huang\textsuperscript{1}}
\\
\textsuperscript{1}Bosch Research,
 \textsuperscript{2}Carnegie Mellon University
\\
}

\begin{document}
\maketitle
\begin{abstract}

Agent skills extend large language model (LLM) agents with reusable, program-like modules that define triggering conditions, procedural logic, and tool interactions. As these skills proliferate in public marketplaces, it is unclear what types are available, how users adopt them, and what risks they pose.
To answer these questions, we conduct a large-scale, data-driven analysis of 40,285 publicly listed skills from a major marketplace. 
Our results show that skill publication tends to occur in short bursts that track shifts in community attention. We also find that skill content is highly concentrated in software engineering workflows, while information retrieval and content creation account for a substantial share of adoption. Beyond content trends, we uncover a pronounced supply-demand imbalance across categories, and we show that most skills remain within typical prompt budgets despite a heavy-tailed length distribution. Finally, we observe strong ecosystem homogeneity, with widespread intent-level redundancy, and we identify non-trivial safety risks, including skills that enable state-changing or system-level actions. Overall, our findings provide a quantitative snapshot of agent skills as an emerging infrastructure layer for agents and inform future work on skill reuse, standardization, and safety-aware design.
\end{abstract}

\section{Introduction}

Large language model (LLM) agents have emerged as a powerful paradigm for addressing complex, multi-step tasks that require reasoning, tool use, and interaction with external environments~\cite{yao2022react,wang2024survey}. Rather than relying on a single prompt-response interaction, agents operate over extended horizons: they interpret user goals, decompose them into sub-tasks, and coordinate actions across diverse tools, data sources, and intermediate states~\cite{shinn2023reflexion,wang2023voyager,sapkota2025ai}. This agentic execution model has enabled a growing range of applications, from software development and data analysis to personal assistance and workflow automation~\cite{wang2024survey}.

\begin{figure}[t]
  \centering
\includegraphics[width=0.98\linewidth]{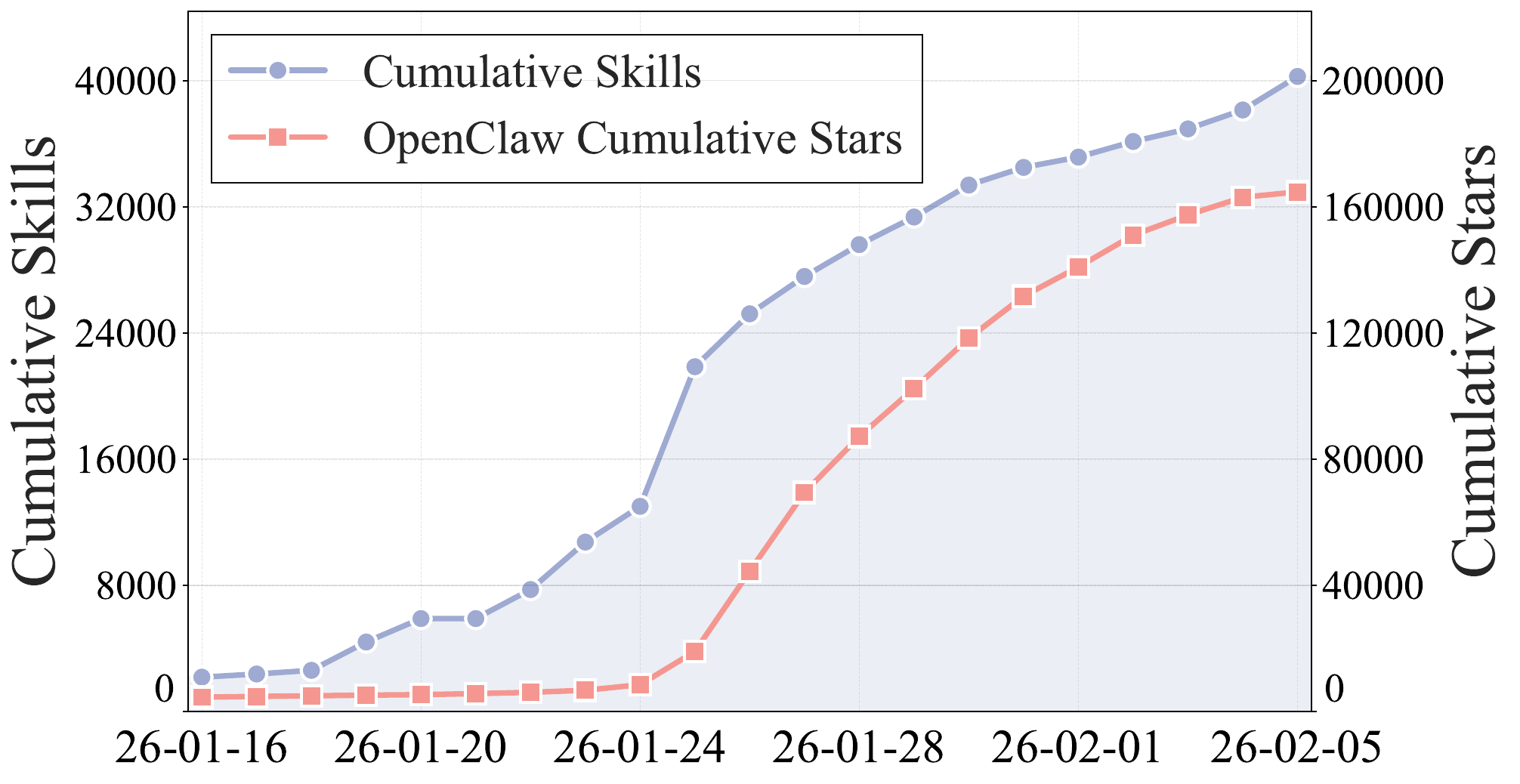}
\vspace{-5pt}
  \caption{\textbf{Growth trend of agent skills}.
According to the well-known agent skills platform \href{https://skill.sh}{skills.sh}, the number of recorded skills experienced rapid growth from mid-January to early February 2026, exceeding 40,000 by early February. During the same period, the popular open-source skills application OpenClaw~\cite{openclaw} saw a sharp surge in GitHub stars, reaching over 25,000 stars in a single day at the end of January, followed by a gradual decline, with the total number of stars exceeding 170k.}
  \vspace{-15pt}
\label{fig:fig1}
\end{figure}

As agent-based systems scale in both complexity and deployment, new challenges arise around reliability, reuse, and maintainability~\cite{liu2026large,cemri2025multi,raheem2025agentic}. Many agent behaviors recur across tasks and users, yet are repeatedly re-specified through prompts or handcrafted control logic~\cite{jin2026agentprimitives,liu2025reuseit}. This has motivated the emergence of agent skills~\cite{openclaw,anthropic2025agentskills}. The ecosystem around these skills has expanded rapidly in recent months, as reflected in Figure~\ref{fig:fig1}. This growth motivates a closer look at what skills exist, how they are adopted, and what risks they may introduce.

\textbf{What are agent skills?}
Agent skills have recently emerged as an abstraction for structuring reusable and scalable agent behaviors~\cite{anthropic2025agentskills,wu2026agent,dive}. The concept has been systematized and popularized by Anthropic in the context of the Claude agent framework~\cite{anthropic2025agentskills}, where skills are sometimes referred to as \emph{Claude Skills} or \emph{Claude agent skills}. An agent skill can be defined as a reusable, plug-and-play module that specifies when the skill should be invoked and how the corresponding subtask should be carried out, typically in a form that can be shared, versioned, and composed.
In practical implementations, a skill often combines lightweight metadata that supports discovery and selection with executable instructions and supporting resources such as files, scripts, and tool configurations~\cite{anthropic2025agentskills}. Appendix~\ref{sec:skill_detail} provides a concrete example of skill structure.

\textbf{Why do agents need skills?}
Agent-based systems often revisit similar subtasks, such as data retrieval, information extraction, code modification, etc.
Without explicit abstractions, these recurring behaviors must be specified repeatedly, either through prompts or through ad hoc control logic. This repetition increases prompt overhead, makes behavior brittle under small context changes, and complicates maintenance of shared procedures~\cite{jin2026agentprimitives}.
Agent skills address these issues by packaging reusable behaviors into modular units that capture task knowledge, procedural logic, and tool use patterns. Because skills can be reused and composed, they improve behavioral consistency and reduce prompt complexity, making agent capabilities easier to extend and control. Shared skill libraries also support standardization and continuous refinement across applications.

\textbf{How are agent skills used in practice?}
In real systems, skills are typically organized as self-contained modules that an agent can select and execute during task solving. A common pattern is to load eligible skills into the system prompt as a compact list of metadata with names and descriptions; the agent then selects a skill based on the user request and follows the skill-defined instructions, issuing tool calls as needed. Appendix~\ref{sec:skill_detail} provides a concrete example (Figure~\ref{fig:skill-integration}) that illustrates how an agent uses different skills to solve user questions.

Taken together, these observations suggest that agent skills are growing quickly, but their functions and potential impacts vary widely. This makes it important to characterize what skills exist, how they are used, and what risks they may introduce. In this paper, we present a large-scale, data-driven measurement of the emerging agent skills ecosystem. Using a corpus collected from a public skill platform, we analyze the following aspects:
\begin{itemize}[leftmargin=0.3cm, itemsep=-0.3em, topsep=-0.2em]
    \item \textbf{Growth trends} in Section~\ref{sec:data_growth}: we quantify publication over time and show rapid, bursty growth.
    \item \textbf{Skill length and redundancy} in Section~\ref{sec:len_redundancy}: we measure prompt length and show a heavy-tailed distribution, while most skills stay within typical prompt budgets. We also find that near-duplicate listings are common.
    \item \textbf{Skill usage patterns} in Section~\ref{sec:supply_demand}: we classify skills into a taxonomy with 6 major categories and 20 sub-categories, and we show strong concentration in software engineering workflows. We further identify clear gaps between what is published and what users install.
    \item \textbf{Safety risks} in Section~\ref{sec:safety}: we audit skills and find that most are low risk, while a non-trivial share enables state-changing actions.
\end{itemize}
Our results point to several open problems. First, the supply--demand gaps and high redundancy call for better skill discovery, de-duplication, and quality signals. Second, the presence of action-enabling skills motivates safety-aware design, including clearer permission models, stronger sandboxing, and more transparent risk labeling.


\section{Skill Data and Growth Trends}
\label{sec:data_growth}

In this section, we introduce our skills dataset and summarizes how the agent-skill ecosystem has grown over time. We first describe how we collect skills and their metadata, and then report statistics that capture publication and adoption patterns.

\subsection{Data Collection }
\label{sec:dataset}

We construct our dataset by crawling agent skills listed on the public marketplace \href{https://skill.sh}{\texttt{skills.sh}}. For each skill, we extract a lightweight set of metadata describing the skill, its hosting location, the date when it first appeared on the marketplace, and how widely it has been installed.
Specifically, each record includes the skill \texttt{name}, a \texttt{repository} field that points to the hosted skill file, \texttt{first\_seen}, which records the date when the skill was first uploaded to \href{https://skill.sh}{\texttt{skills.sh}}, and \texttt{installed\_on}, which reports per-platform installation counts across supported platforms.

We store each skill as a SKILL.md and a JSON-like object with the following structure:
\begin{Verbatim}[breaklines=true,breakanywhere=true,fontsize=\small]
{
  "name": [SKILL NAME],
  "repository": [REPOSITORY LINK],
  "first_seen": [DD/MM/YY],
  "installed_on": [
    {"platform": "claude-code", "installs": [NUMBER OF INSTALL]},
    {"platform": "codex", "installs": [NUMBER OF INSTALL]},
    ...
  ]
}
\end{Verbatim}
\vspace{-5pt}
We finalize data collection on February 5, 2026, yielding \textbf{40,285} skill metadata records. Unless otherwise stated, this snapshot underlies all analyses in Sections~\ref{sec:trend}--\ref{sec:safety}.
All measurements are based solely on publicly accessible content; we avoid sensitive attributions about individual creators and report results only in aggregate.

\subsection{Skill Growth Trends} 
\label{sec:trend}

We study skill growth trends using the \texttt{first\_seen} field in Section~\ref{sec:dataset}, which records when a skill first appears on \href{https://skill.sh}{\texttt{skills.sh}}.
To capture a parallel signal of community attention, we also track the popularity of OpenClaw~\cite{openclaw} by querying its GitHub star history via the GitHub GraphQL API (\texttt{api.github.com/graphql}).
Figure~\ref{fig:fig1} plots the cumulative number of listed skills and the cumulative number of OpenClaw stars.

\paragraph{Growth is rapid and bursty.}
The marketplace grows from 2,179 skills on January 16, 2026 to 40,285 skills on February 5, 2026, a net increase of 38,106 skills in 20 days.
This corresponds to an 18.5$\times$ increase and an average multiplicative growth rate of about 15.7\% per day.
Despite a mean inflow of 1,918 skills per day, arrivals are concentrated in short spikes.
The largest spike occurs on January 25, 2026, when 8,857 skills are added in a single day.
This accounts for 23.2\% of all new skills in the window.
At the weekly level, the week centered on January 25 contributes 19,259 skills, or 47.8\% of the full snapshot.

\paragraph{Growth aligns with an application-level popularity signal.}
OpenClaw exhibits a concurrent popularity shock.
Daily new stars rise from the hundreds in mid-January to 10,543 on January 25, then peak at 25,432 on January 26, which is 2.4$\times$ the previous day.
After the peak, daily gains decline into early February.
For example, OpenClaw gains 1,718 new stars on February 5, which coincides with the slowdown in new listings.
Taken together, the synchronized spikes suggest a shared driver.
A wave of public attention likely encouraged both skill publication and exploration of skill-based tooling.
While GitHub stars are an imperfect proxy for real usage, their synchronized dynamics with marketplace listings provide evidence that rapid supply growth coincides with strong community interest.

\begin{figure}[t]
  \centering
\includegraphics[width=0.98\linewidth]{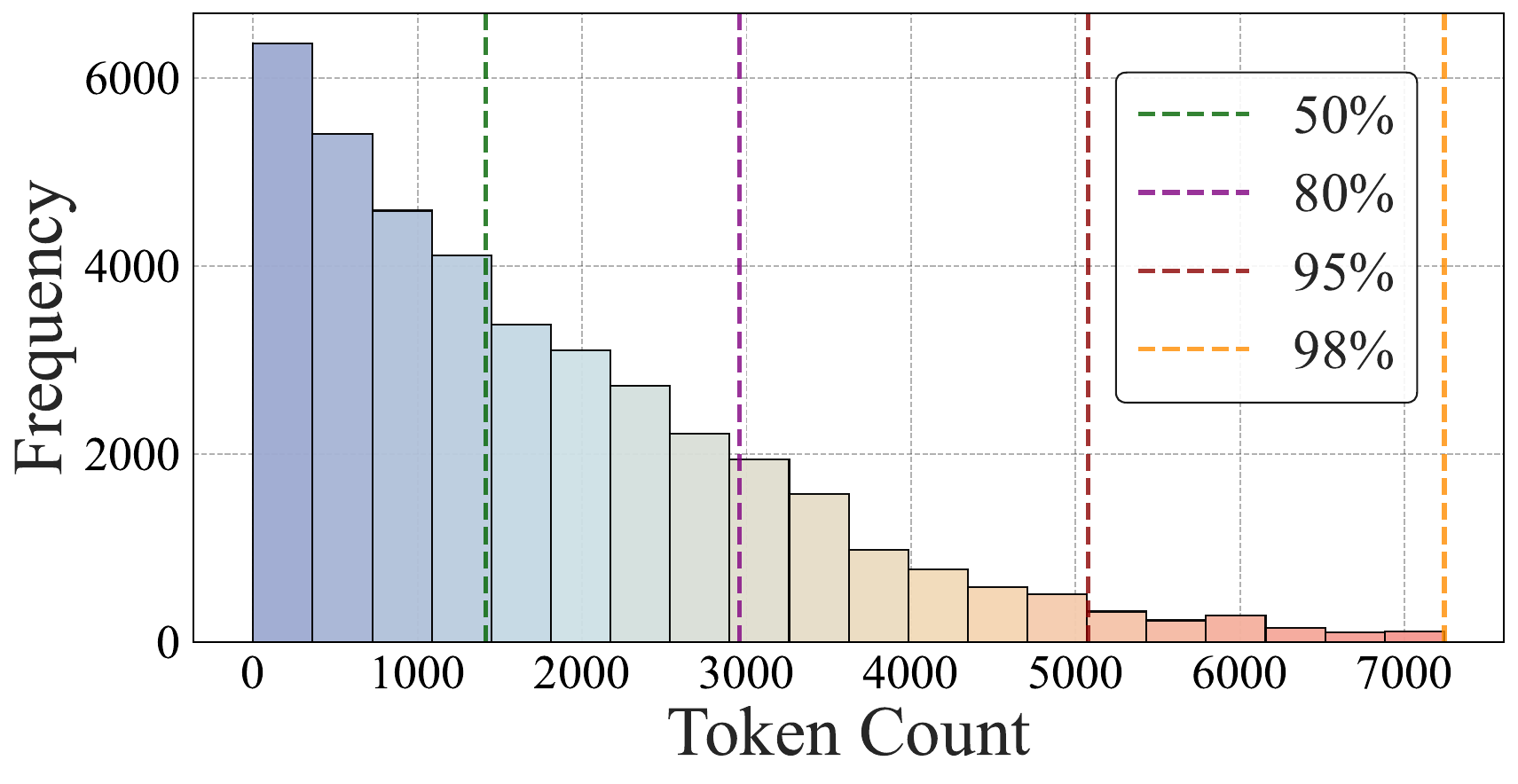}
\vspace{-5pt}
  \caption{\textbf{Token-count distribution of agent skills.}
The distribution is heavy-tailed: the median skill contains 1,414 tokens (mean: 1,895).
90\% of skills are no longer than 3,935 tokens and 99\% are no longer than 9,253 tokens.
A small fraction are exceptionally long, with a maximum of 116,239 tokens.}
\label{fig:length}
\end{figure}

\section{Skill Length and Redundancy}
\label{sec:len_redundancy}

This section summarizes skill content and usage at scale. We first analyze skill length, and then report intent-level redundancy.

\subsection{Skill Length Characteristics} 
\label{sec:length}

We measure skill length by tokenizing each SKILL.md and counting tokens.
For every skill record in Section~\ref{sec:dataset}, we consistently use \texttt{tiktoken}~\cite{tiktoken} with the \texttt{o200k\_base} encoding. This approach allows token counts to serve as a straightforward proxy for prompt budget and inference cost.

\paragraph{Typical skills are short.}
Figure~\ref{fig:length} shows a pronounced heavy-tailed distribution, but most skills are still compact.
The median length is 1,414 tokens and the mean is 1,895 tokens, indicating that a typical skill fits comfortably alongside planning context and tool schemas.
Across quantiles, length is rarely a binding constraint: 80\% and 95\% of skills are within 2,955 and 5,077 tokens, and 90\% and 99\% remain below 3,935 and 9,253 tokens.
Meanwhile, the standard deviation of 2,025 tokens highlights substantial dispersion, even though the central mass is short.

\paragraph{A few skills are long due to the inclusion of multiple components.}
This dispersion is driven by a small set of extreme outliers.
The top 1\% of skills exceed 9,253 tokens, and the maximum reaches 116,239 tokens, which can consume prompt budgets and hinder reliable selection and auditing when loaded in full.
Manual inspection suggests that many of these long skills consolidate multiple components into a single file, including extended documentation, large code blocks, and reusable template collections.
This pattern indicates that, while most skills support direct in-context use, a minority behave like libraries and may benefit from modularization or retrieval-based loading so only the relevant portions are brought into context.

\begin{figure}[t]
  \centering
\includegraphics[width=0.98\linewidth]{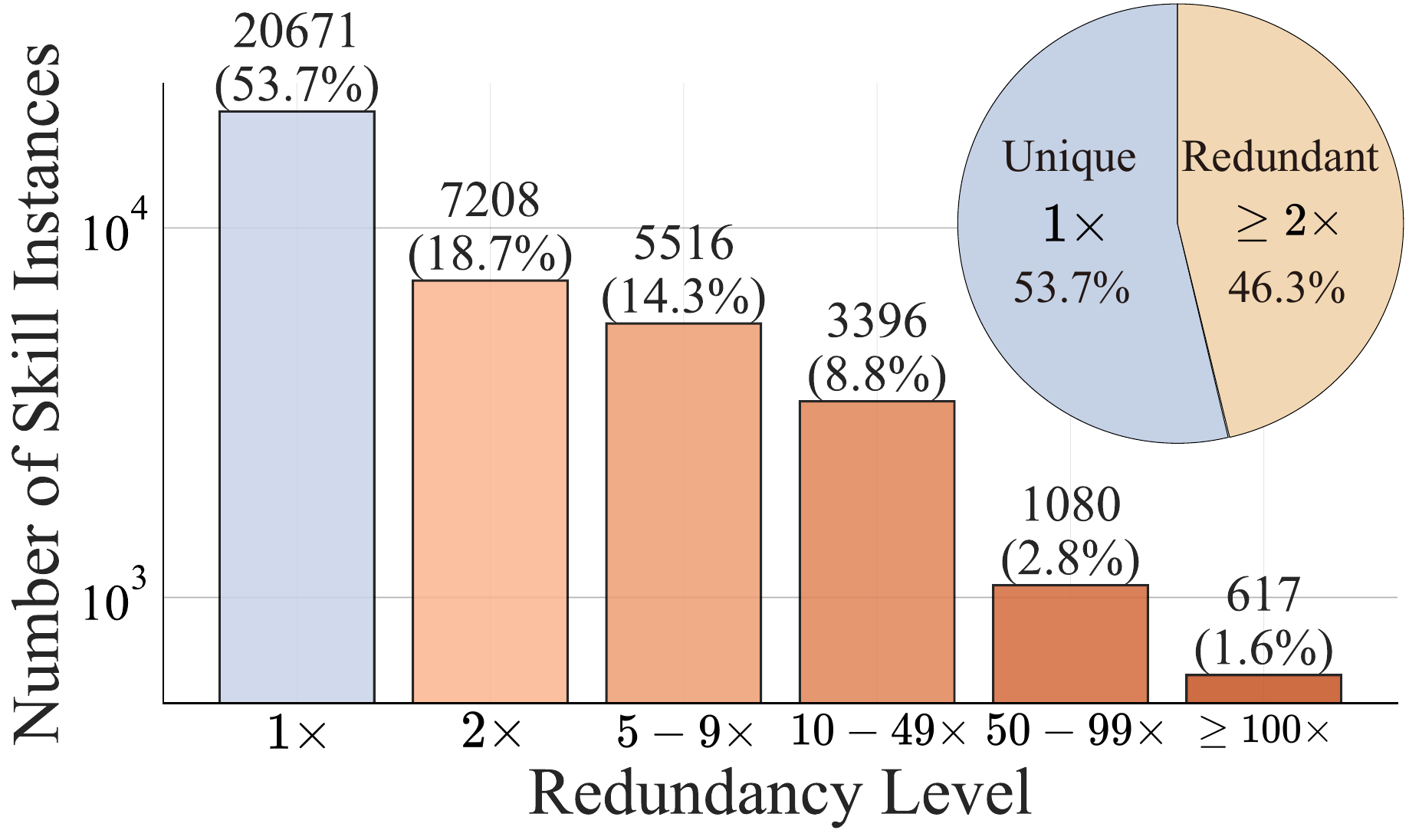}
\vspace{-5pt}
  \caption{\textbf{Name based redundancy distribution.} Skills are grouped by normalized names using case insensitive matching after removing special characters. We report the fraction of skills that appear $n$ times, denoted as $n\times$. Skills that appear once account for 53.7\%, while skills that appear more than once account for 46.3\%. The names of the 30 most redundant skills under this metric are listed in Appendix Figure~\ref{fig:top_redundant}.}
  \vspace{-15pt}
\label{fig:redundancy}
\end{figure}
\begin{figure}[t]
  \centering
\includegraphics[width=0.99\linewidth]{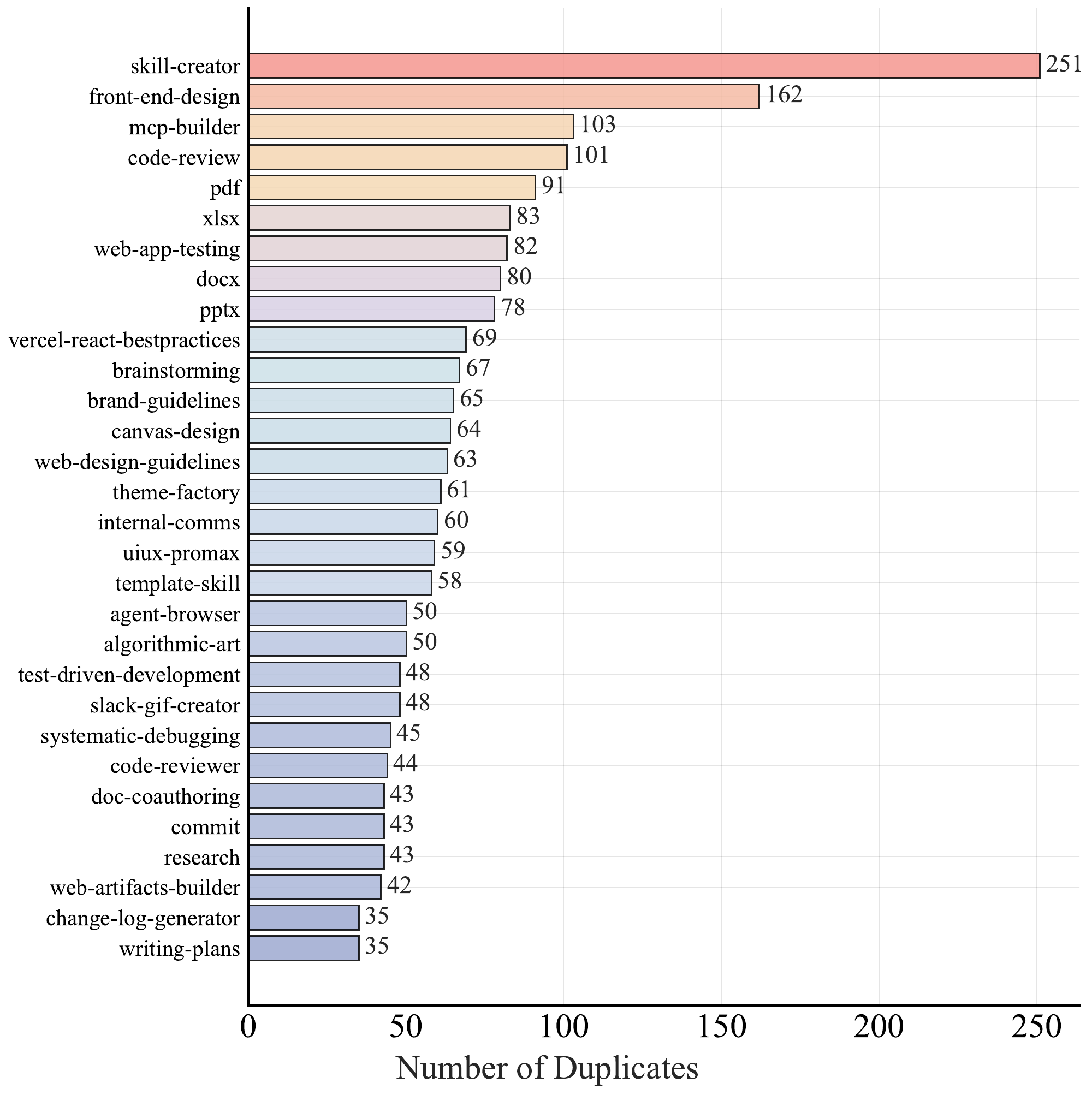}
  \caption{\textbf{Top 30 redundant skills by name based matching.} We rank skills by the number of listings that share the same normalized name, using the exact matching procedure in Section~\ref{sec:redundancy}. This figure lists the 30 most frequently repeated skill names and their repetition counts.}
\label{fig:top_redundant}
\end{figure}

\begin{table*}[t]
\centering  
\resizebox{0.99\textwidth}{!}{
\begin{tabular}{llrccc}
\toprule
\rowcolor{gray!15}
\textbf{Major Category} & \textbf{Sub-Category} & \textbf{\# Skills} & \textbf{\% of Total} & \textbf{Avg Tokens} & \textbf{Avg Downloads} \\
\midrule
Software Engineering & Code Generation & 5,743 & 14.3\% & 2004 & 235 \\
 & Debug \& Analysis & 5,319 & 13.2\% & 1772 & 103 \\
 & Version Control & 1,275 & 3.2\% & 1403 & 71 \\
 & Infrastructure & 9,664 & 24.0\% & 1995 & 114 \\
\midrule
Information Retrieval & Web Search & 567 & 1.4\% & 1517 & 1268 \\
 & Academic Search & 1,083 & 2.7\% & 2100 & 73 \\
 & Live Data Streams & 277 & 0.7\% & 1514 & 48 \\
\midrule
Productivity Tools & Team Communication & 698 & 1.7\% & 1458 & 196 \\
 & Document Systems & 1,579 & 3.9\% & 1981 & 125 \\
 & Task Management & 2,275 & 5.6\% & 1656 & 106 \\
\midrule
Data \& Analytics & Data Processing & 3,179 & 7.9\% & 2134 & 93 \\
 & Math \& Calculation & 368 & 0.9\% & 2028 & 147 \\
 & Data Visualization & 736 & 1.8\% & 2322 & 108 \\
\midrule
Content Creation & Image Generation & 1,201 & 3.0\% & 2145 & 214 \\
 & Text Generation & 2,212 & 5.5\% & 1977 & 178 \\
 & Audio \& Video & 1,466 & 3.6\% & 1744 & 266 \\
\midrule
Utilities \& Other & Local File Control & 255 & 0.6\% & 1420 & 42 \\
 & Command Execution & 337 & 0.8\% & 1541 & 70 \\
 & Memory \& Cognition & 929 & 2.3\% & 1475 & 54 \\
 & Other Utilities & 1,125 & 2.8\% & 1684 & 135 \\
\bottomrule
\end{tabular}
}
\caption{\textbf{Functional taxonomy and category-level statistics of agent skills.} Skills are organized into 6 major categories and 20 sub-categories. For each sub-category, we report its size (\# skills and \% of corpus), the average skill length in tokens, and the mean downloads/installs.}
\label{tb:gpt_stats}
\end{table*}

\begin{figure*}[t]
  \centering
\includegraphics[width=0.99\linewidth]{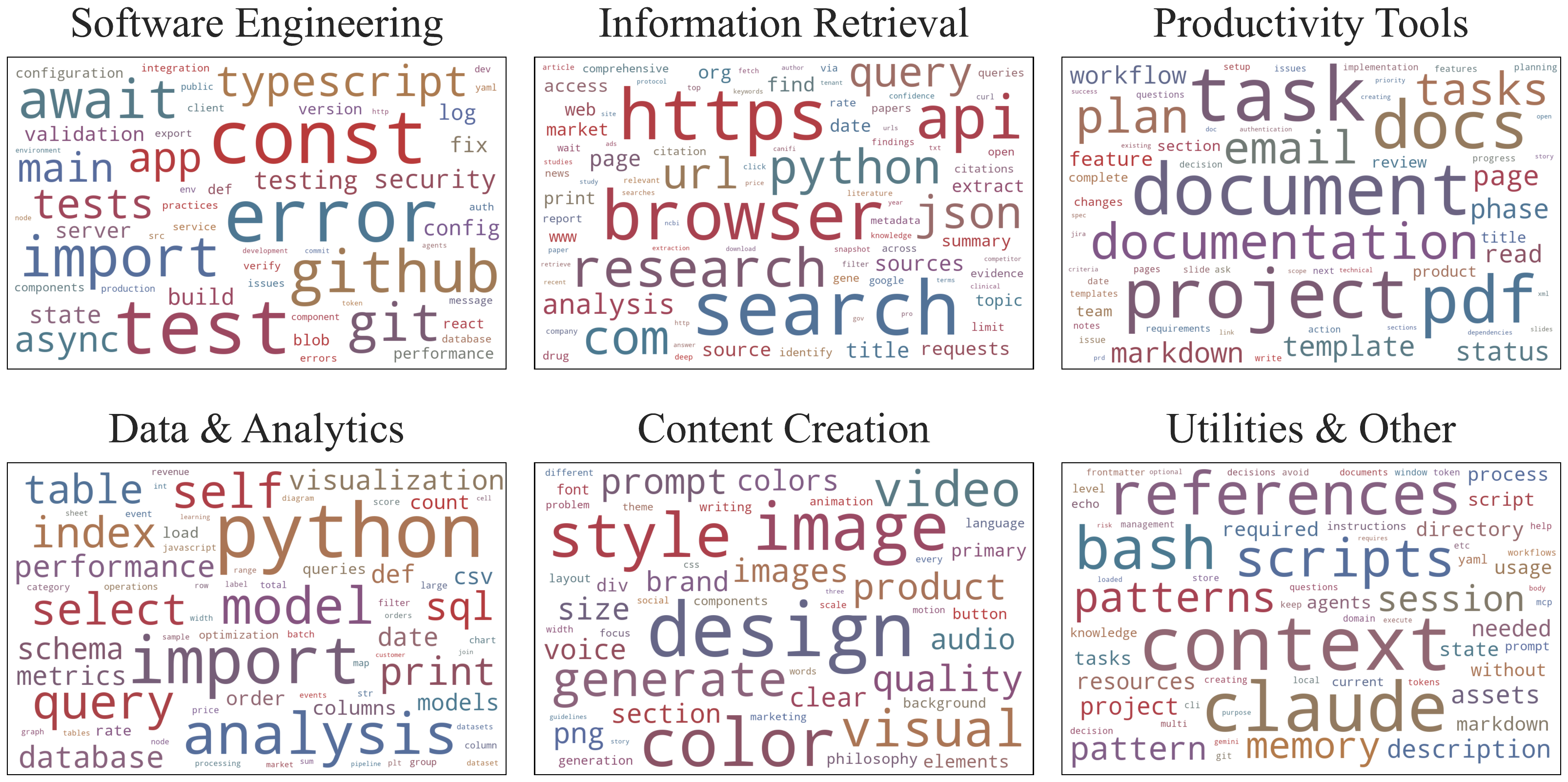}
  \caption{\textbf{Word clouds of skill names by major category.} For each major category in Section~\ref{sec:content}, we show the most frequent terms derived from the skill document. Words are retained only if their frequency in the target category exceeds $1.5$ times the average of the remaining five categories. Font size is proportional to within-category frequency, highlighting common topics and recurring workflow motifs.}
\label{fig:framework}
\end{figure*}

\subsection{Intent-level Redundancy Analysis} 
\label{sec:redundancy}

As the agent skills ecosystem grows, identical user intents are frequently published more than once, either by independent developers or through template-driven generation.
Moderate repetition can be beneficial when it produces meaningfully improved variants, such as stronger safety checks, broader tool coverage, or clearer documentation.
However, when most copies differ only in wording, marketplace volume rises without a corresponding increase in capability diversity.
We therefore quantify redundancy as the frequency with which the same intent is re-listed in the corpus.

\paragraph{Redundancy measuring.}
We estimate redundancy with two signals.
First, we apply name-based exact matching.
We lowercase skill names, remove special characters, and group skills that share the same normalized name.
Second, we apply semantic matching.
We encode \texttt{Name:[NAME] + Description:[DESCRIPTION]} with \texttt{BAAI/bge-m3} \cite{chen2024bge}, and analyze nearest neighbor similarity together with t-SNE visualizations.
In this marketplace, descriptions are often short, noisy, and template derived, so semantically similar embeddings do not reliably separate true duplicates from loosely related skills.
For this reason, we use strict name matching for the main results and report the embedding analysis in Appendix \ref{appendix:redun}.

\paragraph{Nearly half of listings are duplicates.}
Figure~\ref{fig:redundancy} shows that unique entries only slightly outnumber repeated ones.
Under strict exact matching, 53.7\% of skills appear once, while 46.3\% share a normalized name with at least one other listing.
Duplication is also concentrated.
Pairs are common, with 2$\times$ groups contributing 18.7\% of the corpus.
Higher multiplicities still account for a nontrivial share: 5$\times$ to 9$\times$ groups contribute 14.3\%, and 10$\times$ to 49$\times$ groups contribute 8.8\%.
A small number of names appear more than 100 times, which is consistent with repeated reposting or automated publication from shared templates. For concreteness, we provide the 30 most redundant skill names in Figure~\ref{fig:top_redundant}.

\paragraph{Implications for discovery and maintenance.}
High redundancy increases user search costs and fragments feedback and adoption signals across near identical listings, which makes it harder for high quality implementations to become clear defaults.
It also indicates that developer effort is often spent re-packaging common workflows rather than expanding coverage into less served tasks.
This motivates platform mechanisms that encourage reuse and differentiation, including clearer canonical skills, more explicit versioning, and modular templates that reduce incentives to publish superficial copies.

\section{Skill Usage Patterns} 
\label{sec:supply_demand}
In this section, we study what skills do and what users actually install. We summarize skill functionality with a taxonomy to support a corpus-level comparison of skills' publication with adoption.

\subsection{Taxonomy and Classification}
We define a two-level taxonomy with 6 major categories and 20 sub-categories (Table~\ref{tb:gpt_stats}).
The taxonomy covers end-to-end agent workflows and separates common intents that differ in practice, such as Code Generation and Debug \& Analysis.
For each sub-category, we report the number of skills, the mean token length, and the mean downloads or installs.
Figure~\ref{fig:framework} provides qualitative signals by showing category-wise word clouds from frequent words in full skill markdown documents. We retained words only if their frequency in the target category exceeded $1.5$ times the average of the remaining five categories. 
Marketplace tags are sparse and inconsistent, so we label skills with Qwen2.5-32B-Instruct \cite{qwen2025qwen25technicalreport}.
Given a skill's \texttt{name} and \texttt{description}, the model selects one sub-category and returns a strict JSON record.
Appendix~\ref{app:classifier_prompt} lists the taxonomy definitions and the prompt template.

\begin{figure}[t]
  \centering
\includegraphics[width=0.98\linewidth]{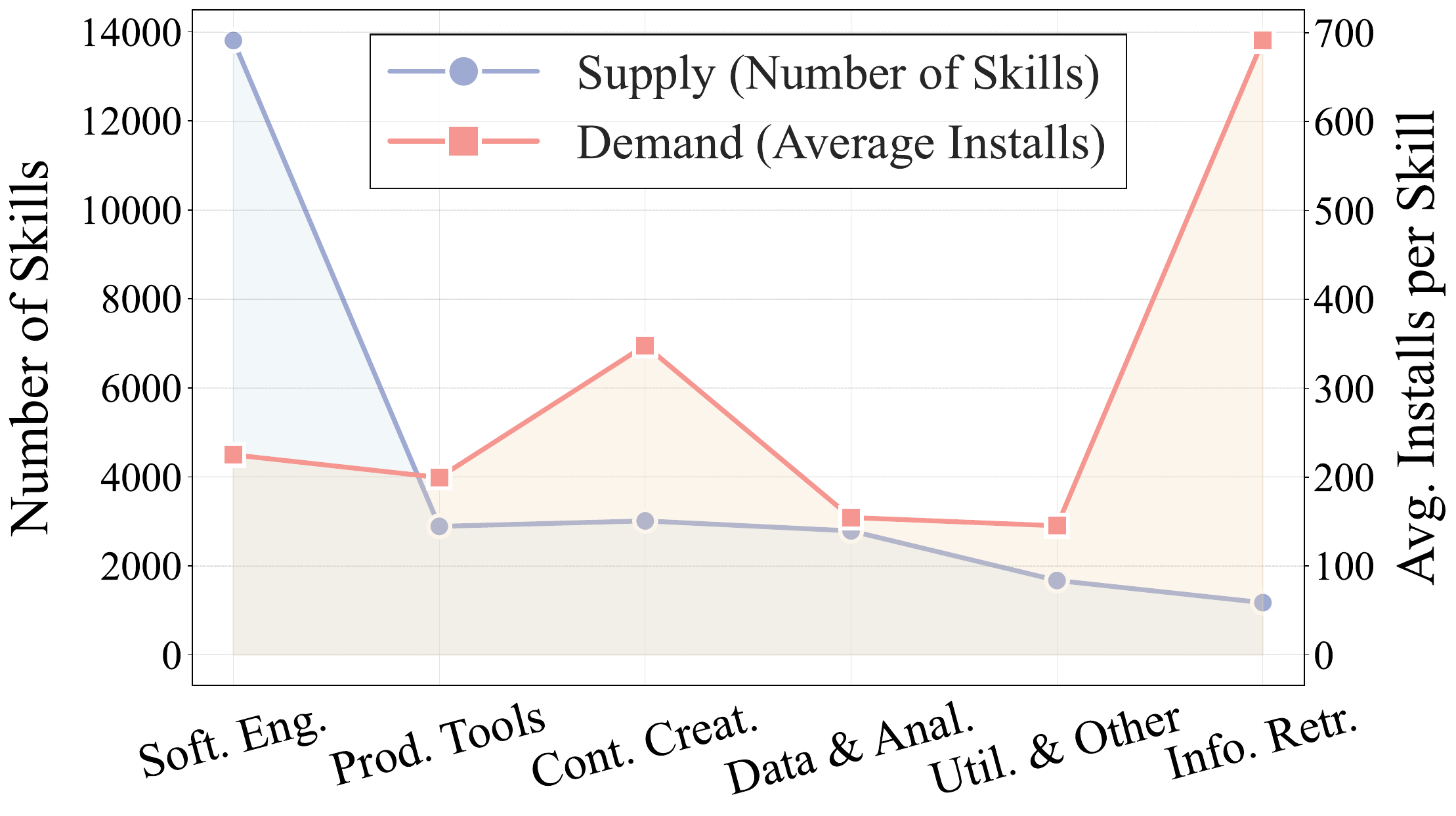}
  \caption{\textbf{Supply--demand dynamics by major category.} Supply denotes the number of de-duplicated skills after redundancy filtering (Section~\ref{sec:redundancy}). Demand denotes the average installs per skill and is used as a coarse proxy for adoption. Most categories are close to balanced, while content creation is demand-heavy, software engineering is supply-heavy, and information retrieval is demand-heavy despite limited supply.}
\label{fig:supply_demand}
\end{figure}

\subsection{Distribution of Supply and Demand} 
\label{sec:content}


\paragraph{Data-centric skills are longer.}
Token length varies by function.
Data Visualization and Data Processing are the longest sub-categories on average, at 2,322 and 2,134 tokens.
These skills often include multi-step pipelines, configuration blocks, and reusable templates.
By comparison, Version Control and Local File Control are shorter at 1,403 and 1,420 tokens and are typically more procedural.
These differences matter for prompt budgeting and auditing, since a small category can still impose a large review burden if its skills are long.

\paragraph{Software Engineering dominates listings.}
In Table~\ref{tb:gpt_stats}, Software Engineering accounts for 54.7\% of the corpus across Code Generation, Debug \& Analysis, etc.
Infrastructure is the largest sub-category with 9,664 skills, or 24.0\% of all listings.
This suggests that developers often publish skills for environment setup, DevOps automation, tool configuration, and deployment.

\paragraph{Adoption concentrates on a few general skills.}
Adoption follows a different pattern from publication.
Web Search has the highest mean downloads at 1,268, but it represents only 1.4\% of listings.
This indicates that a small number of retrieval connectors are reused widely.
Content creation also shows high mean installs, with Audio \& Video at 266 and Image Generation at 214.
Code Generation remains among the most installed at 235.
In contrast, utility-focused skills have lower mean installs, including Local File Control at 42 and Memory \& Cognition at 54.
This may reflect narrower use cases, higher perceived risk, or overlap with built-in agent capabilities.
We analyze this publication adoption gap further in Section~\ref{sec:supply_demand}.

Overall, the taxonomy provides a compact map of the ecosystem.
It distinguishes what developers build from what users reuse and clarifies how content complexity varies across functions.

\subsection{Supply--Demand Dynamics}

Ecosystem growth alone does not show whether developers publish the capabilities that users adopt most.
We therefore compare supply and demand across functional categories. Following Section~\ref{sec:redundancy}, we de-duplicate skills to reduce the impact of near-identical reposts and template variants.
We define supply as the number of de-duplicated skills in a category, and define demand as the average installs per skill and use it as a coarse proxy for adoption.
\begin{figure}[t]
  \centering
\includegraphics[width=0.99\linewidth]{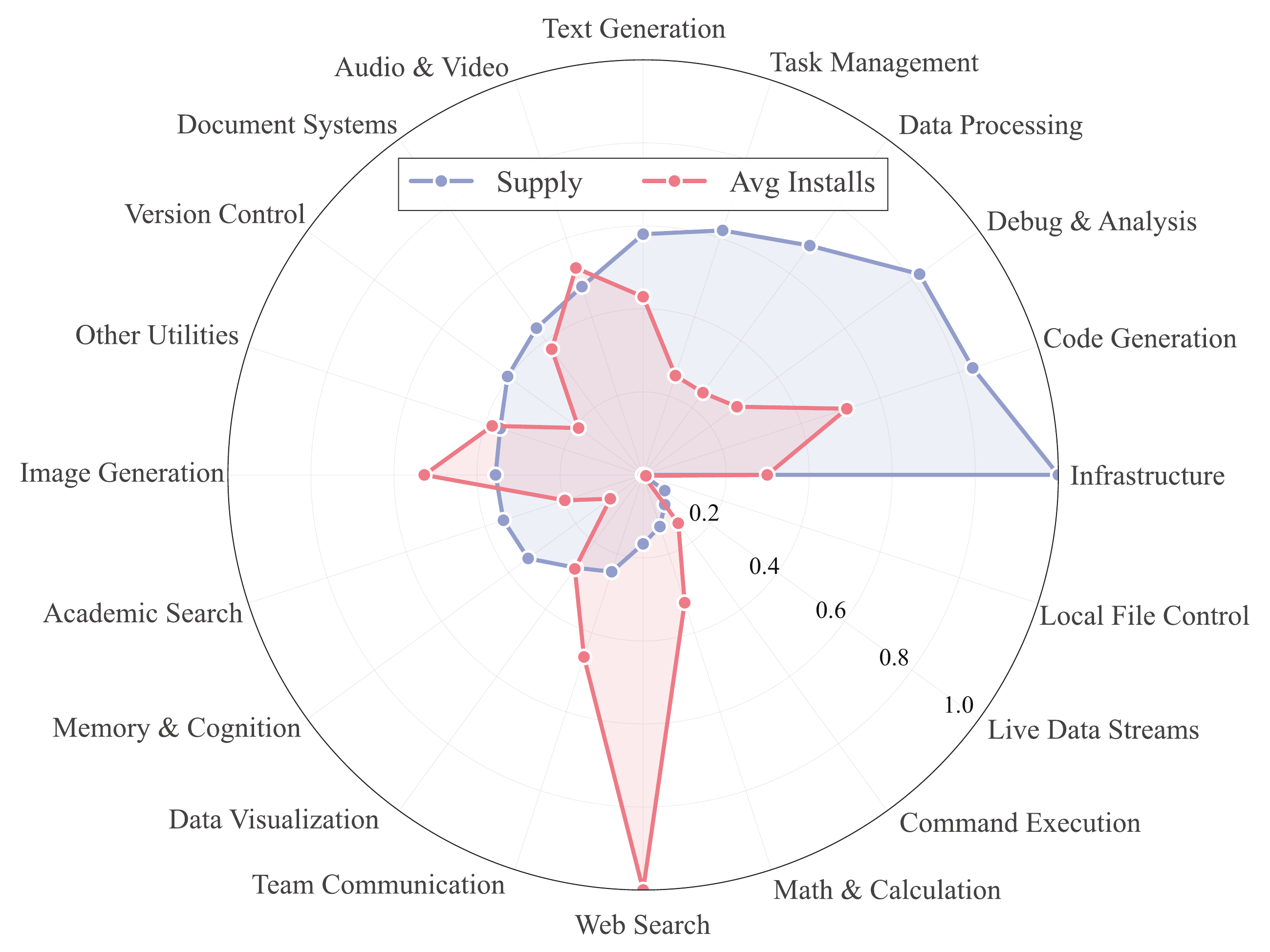}
  \caption{\textbf{Supply--demand dynamics by sub-category.} Supply denotes the number of de-duplicated skills in each of the 20 sub-categories, and demand denotes the average installs per skill. The figure highlights where publication volume and adoption diverge at a finer granularity than Figure~\ref{fig:supply_demand}.}
\label{fig:radar}
\end{figure}

\paragraph{Broad alignment with systematic deviations.}
Figure~\ref{fig:supply_demand} summarizes the results.
In most categories, supply and demand move in the same direction, suggesting that publication broadly tracks user interest. However, we observe three consistent gaps.
First, content creation is demand-heavy.
Users repeatedly reuse writing and media workflows, even when the number of listings is modest.
Second, software engineering is supply-heavy.
Skills that wrap coding, testing, and repository routines are easy to produce and share, which increases overlap and spreads installs across close substitutes.
Third, information retrieval \cite{zheng2025skillweaver, wu2025generativesearchengineslike} is demand-heavy despite limited supply.
A small number of reliable retrieval skills can attract many installs because web and database access is useful for everyday tasks.
However, publishing these skills is costly because it requires stable connectors, careful query design, and ongoing maintenance as external interfaces and rate limits change.

Figure~\ref{fig:radar} further provides the same comparison for all 20 sub-categories.
Together, these results highlight where canonical implementations, stronger tool integration, and maintenance incentives may reduce supply-demand mismatches.

\section{Risk and Safety Assessment}
\label{sec:safety}

Skills are executable procedures that interact with external services, local environments, and user context.
Relative to prompt-only interactions, they expand the harm surface because a skill may access sensitive data or trigger real-world side effects~\cite{wu2026agent,schmotz2025agent,liu2026agent}.
We therefore quantify how often published skills enable privacy-sensitive reads, state-changing actions, or critical capabilities such as arbitrary command execution.

\paragraph{LLM-based auditing protocol.}
We audit each skill with Qwen2.5-32B-Instruct using the rubric in Appendix~\ref{app:security_audit_prompt}.
The model receives the skill name, description, and full SKILL.md content, and assigns exactly one risk level from \texttt{L0} to \texttt{L3} under a worst-case interpretation, where \texttt{L0} is safe, \texttt{L1} is privacy risk, \texttt{L2} is moderate risk, and \texttt{L3} is critical risk.
To support automatic aggregation, we require a strict JSON response:
\begin{Verbatim}[breaklines=true,breakanywhere=true,fontsize=\small]
{"skill_name": "{{SKILL_NAME}}", "risk_level": "L0" | "L1" | "L2" | "L3", "reasoning": "A concise sentence explaining the specific risk factor."}
\end{Verbatim}
We report results by the six major categories from Section~\ref{sec:content}.
For a finer breakdown across all 20 sub-categories, see Appendix Figure~\ref{fig:risk_20}.
\begin{figure}[t]
  \centering
\includegraphics[width=0.98\linewidth]{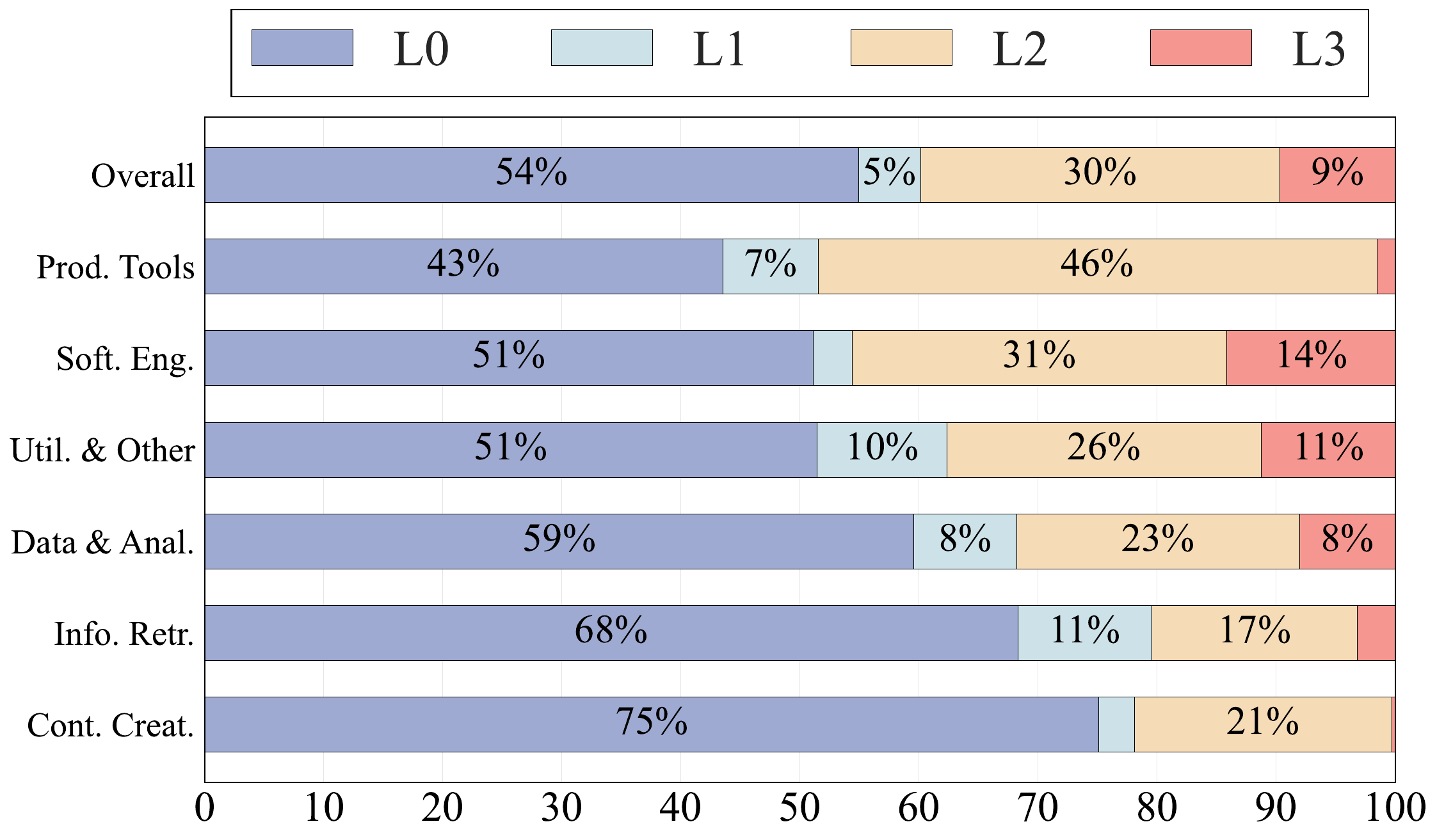}
  \caption{\textbf{Risk level distribution} overall and by major category, where \texttt{L0} is the lowest risk and \texttt{L3} is the highest, audited with Qwen2.5-32B-Instruct using Appendix~\ref{app:security_audit_prompt}.}
  \vspace{-15pt}
\label{fig:risk}
\end{figure}
\paragraph{Overall distribution.}
Figure~\ref{fig:risk} shows that low-risk skills dominate, but action-enabling skills are widespread.
Overall, 54\% are \texttt{L0}, 5\% are \texttt{L1}, 30\% are \texttt{L2}, and 9\% are \texttt{L3}.
Thus, nearly two fifths of the marketplace can access sensitive context or perform writes and actions, and a nontrivial share exposes critical capabilities.

\begin{figure}[t]
  \centering
\includegraphics[width=0.99\linewidth]{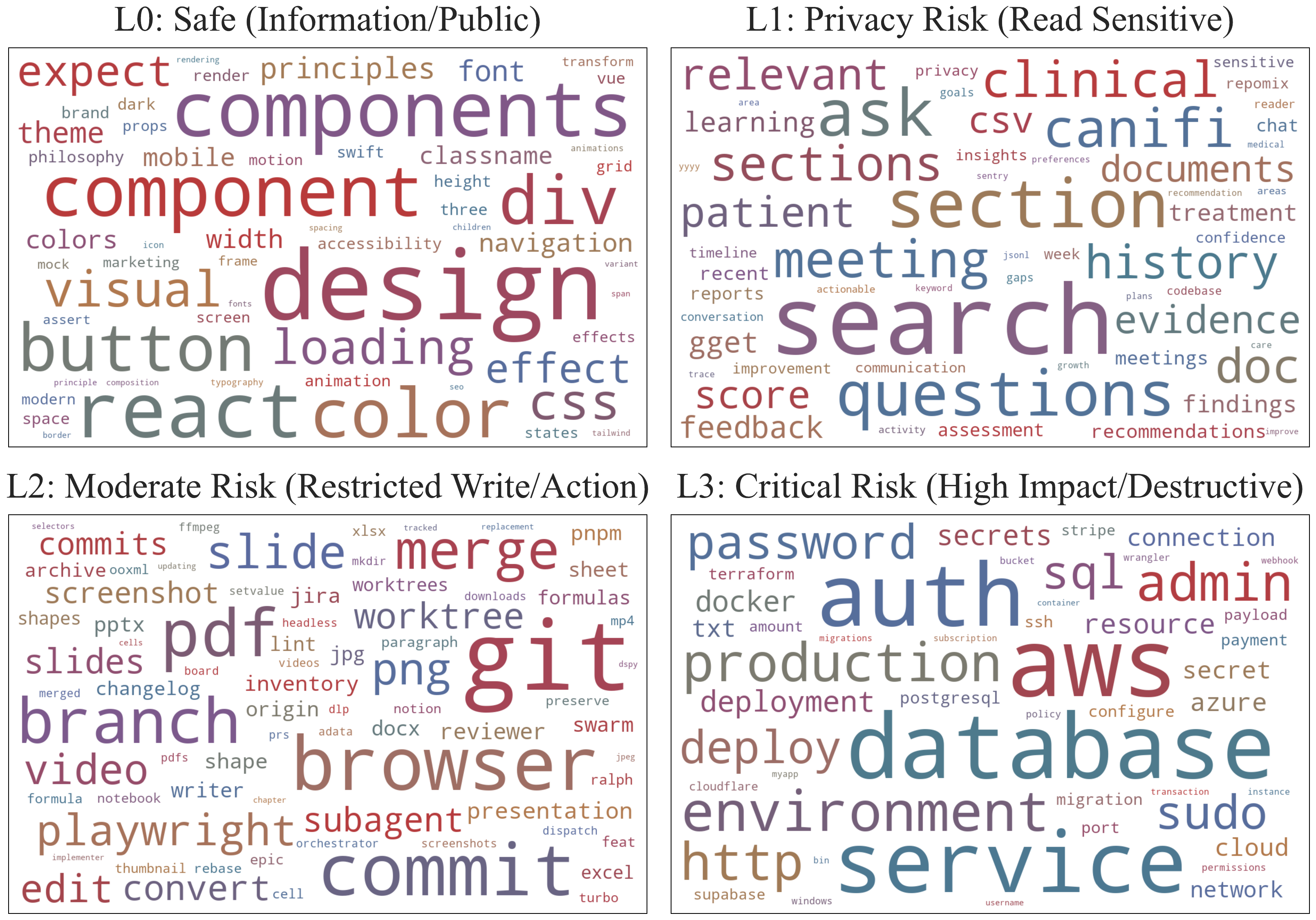}
  \caption{\textbf{Risk level wordcloud.} L0 lists safe design words like "visual" and "component." L1 focuses on private data with terms like "meeting" and "history." L2 shows action words like "git" and "merge." L3 highlights critical system and security terms such as "sudo," "admin," and "password."}
\label{fig:risk_wordcloud}
\end{figure}

\paragraph{Category-level patterns.}
Risk concentrates in categories that connect the model to external systems.
Content Creation is the safest category, with 75\% \texttt{L0} and only a small \texttt{L3} share, which matches workflows whose outputs are mainly drafts or media artifacts.
Information Retrieval is largely read oriented, with 68\% \texttt{L0} and the largest \texttt{L1} share at 11\%, often because connectors rely on user-specific tokens or private sources.
Productivity Tools is dominated by \texttt{L2} at 46\%, reflecting common actions such as creating, editing, and sending emails, messages, calendar entries, and documents.
Software Engineering has the highest \texttt{L3} fraction at 14\%, consistent with skills that manage environments, run commands, or manipulate repositories.
Utilities \& Other also shows elevated \texttt{L1} at 10\% and \texttt{L3} at 11\%, driven by local file operations and command execution utilities.
Data \& Analytics falls between these extremes, with 59\% \texttt{L0} and 23\% \texttt{L2}, consistent with ETL-style pipelines that may write intermediate outputs. Figure \ref{fig:risk_wordcloud} further presents a detailed word cloud analysis across different risk levels from L0 to L3. Words are retained only if their frequency in the target level exceeds $1.5$ times the average of the remaining three levels.

Overall, the most severe cases are less about content generation and more about enabling external side effects.
These results motivate least-privilege tool design and additional safeguards for high-risk operations.

\section{Potential Directions}
In this paper, our measurements suggest that the agent skill ecosystem is at an inflection point.
Progress now depends on improving quality, reducing overhead, and managing risk at scale.

\textbf{First}, rapid growth of Agent Skills relies on community contribution, but long term value depends on high quality, non redundant skills.
Given pervasive intent level duplication, future work should pair semantic de duplication with quality signals including documentation, execution reliability, maintenance, and usage.
A practical goal is convergence to a small set of canonical skills per intent, so developers extend capabilities instead of re packaging the same workflows.

\textbf{Second}, the heavy tailed length distribution of skills' tokens means a small fraction of skills can dominate prompt budgets.
Future systems should support selective loading and modularization, retrieving only the steps, parameters, and tool schemas needed for the current subgoal.
Summarization, pruning unused branches, and instruction compression can further reduce overhead while preserving faithful execution.

\textbf{Third}, our observed supply-demand gaps in Agent Skills indicate that publication effort does not always track user adoption.
In particular, information retrieval attracts high usage but remains costly to build and maintain, while many software engineering skills compete as close substitutes.
Future Skills platforms can use demand signals to guide authoring tools, incentives, and review effort, and can support demand driven synthesis that adapts existing skills to new connectors, domains, and user constraints.

\textbf{Finally}, the presence of high-risk skills calls for the establishment of proactive security protocols. Current frameworks lack fine-grained control over state-changing actions. Future research should implement standardized sandboxing environments that enforce the principle of least privilege. Such protocols would allow agents to perform complex tasks while protecting the host system from unauthorized or malicious operations.

\section{Conclusion}
\label{sec:conclusion}
In this paper, we conducted a large-scale, data-driven measurement of the agent skills ecosystem, analyzing over 40,000 publicly listed skills.
Across production, adoption, redundancy, and safety auditing, our results provide a quantitative snapshot of skills as an emerging abstraction for extending large language model agents.
Overall, the ecosystem is expanding quickly but unevenly: supply is dominated by software engineering skills, intent-level redundancy is pervasive, and adoption concentrates on a smaller set of high-demand capabilities, notably information retrieval and content creation.
From a safety perspective, although most skills appear low risk, a non-trivial subset enables state-changing or system-level actions, underscoring the need for safety-aware skill design and review.
Taken together, these findings position agent skills as a measurable infrastructure layer for agent systems and motivate future work on standardization, reuse, and governance.

\clearpage

\section*{Limitation}
\label{sec:Limitation}
Our study has two main limitations. First, our measurements are derived from a single snapshot of one public marketplace collected around early February 2026. As platform policies, ranking algorithms, and community composition evolve, both the supply and the demand for skills may change, and the bursty growth patterns we report may not persist. Second, we operationalize adoption using publicly visible signals rather than verified executions inside deployed agents. These signals can be affected by interface changes, caching, and social dynamics such as coordinated promotion, and they may under represent private usage that occurs through enterprise deployments or custom agent stacks.

\bibliography{custom}
\clearpage
\appendix

\section{Skill Structure and Integration in Agents}
\label{sec:skill_detail}

Skills provide a modular and expressive mechanism for extending agent capabilities. By encapsulating triggering conditions, procedural logic, and tool interactions, skills enable agents to compose complex behaviors dynamically at runtime, while keeping the core agent architecture lightweight and unchanged.
\begin{figure}[t]
  \centering
  \vspace{-5pt}
\includegraphics[width=0.99\linewidth]{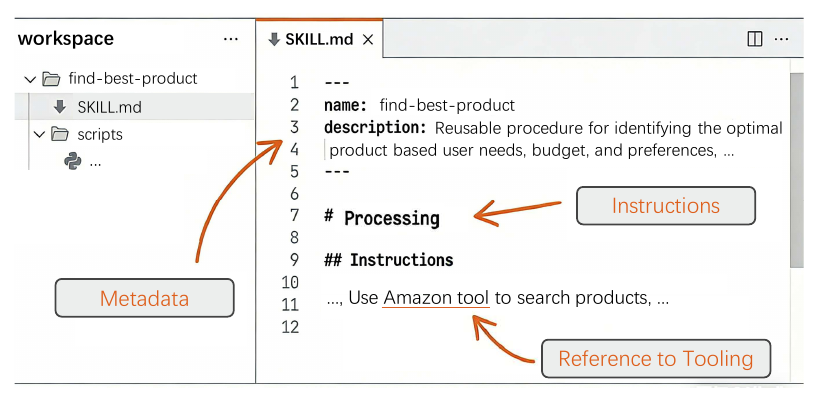}
\vspace{-8pt}
\caption{Internal structure of a typical agent skill, illustrated using the \texttt{find-best-product} skill. The \texttt{SKILL.md} file begins with YAML metadata that specifies the skill name and description, which are used for skill discovery and selection. The subsequent Markdown sections define the procedural workflow and detailed execution instructions, including references to external tools such as product search APIs. This structure enables lightweight semantic matching at discovery time while supporting complex, tool-integrated execution when the skill is invoked.}
  \vspace{-8pt}
\label{fig:skill-structure}
\end{figure}
\begin{figure}[t]
  \centering
  \vspace{-5pt}
\includegraphics[width=0.99\linewidth]{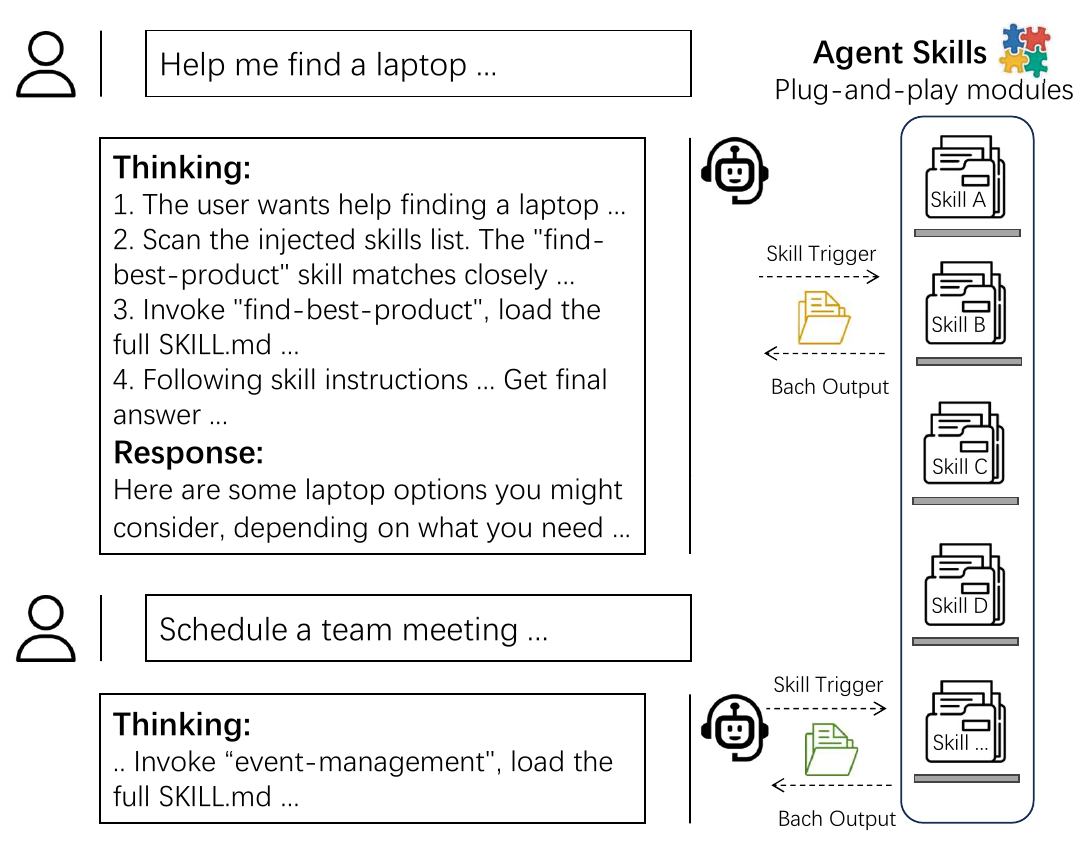}
\vspace{-8pt}
\caption{Dynamic skill integration during agent execution. Given a user request, the agent reasons over the injected skill index and selects the most relevant skill based on semantic matching. The full skill specification is then loaded, and the agent follows the skill's instructions, potentially invoking external tools and producing intermediate outputs. Different user intents trigger different skills, enabling modular, reusable, and context-aware agent behavior through conditional skill selection and state-driven transitions.}
  \vspace{-8pt}
\label{fig:skill-integration}
\end{figure}
\subsection{Skill Structure}

Skills act as first-class capability abstractions in an agent system, allowing agents to solve complex tasks through structured and reusable procedures. Unlike ad hoc prompts or isolated tool calls, each skill is defined as a lightweight, program-like unit with explicit execution semantics. Formally, a skill can be represented as:
\begin{Verbatim}[breaklines=true,breakanywhere=true,fontsize=\small]
Skill = {Metadata, Instructions, Resources},
\end{Verbatim}
where metadata specifies the applicability conditions of the skill, instructions define the procedural steps to be executed, and resources link the skill to external tools, scripts, APIs, or auxiliary artifacts.

In practice, each skill is organized within a dedicated directory containing a \texttt{SKILL.md} file that materializes this abstraction. As shown in Fig.~\ref{fig:skill-structure}, \texttt{SKILL.md} begins with a YAML frontmatter that specifies the skill name and a concise description. This metadata serves as a compact semantic signal for skill discovery and selection. The remainder of the file is written in Markdown and encodes the procedural logic of the skill, including high-level processing stages, decision criteria, and concrete instructions for tool invocation.
This design separates lightweight skill discovery from full procedural execution. During agent initialization, only the metadata is exposed to the model, minimizing prompt overhead. Detailed instructions and associated resources are loaded only when a skill is selected, enabling rich, tool-augmented behavior without inflating the agent's baseline context.

\subsection{Skill Integration in Agents}

Beyond the definition of individual skills, agent intelligence critically depends on how multiple skills are dynamically coordinated during task execution. At the start of a session, the agent is provided with a summarized index of available skills, consisting of their names and descriptions. Given a user request, the agent performs high-level reasoning to infer the underlying goal and decompose it into subgoals. Rather than following a fixed execution pipeline, the agent conditionally selects skills based on the current subgoal and the evolving internal state.

Once a skill is invoked, the agent follows the skill's procedural instructions, potentially interacting with external tools and producing structured intermediate outputs. These outputs are incorporated into the agent's internal state and directly influence subsequent decisions. For instance, the execution of an analysis skill may introduce new constraints or observations, which in turn determine whether the agent invokes a refinement skill, explores alternative strategies, or proceeds to a synthesis step. As a result, skill execution induces a non-linear control flow in which transitions between skills are determined by intermediate state updates rather than pre-defined task graphs. Through iterative reasoning, skill selection, and execution, multiple skills collaboratively contribute to the completion of complex tasks.

Fig.~\ref{fig:skill-integration} illustrates this execution paradigm. The agent alternates between reasoning over the current state, selecting an appropriate skill from the available index, and executing the selected skill to update its state. Importantly, this coordination emerges from model-driven decision-making over skill abstractions, without relying on hard-coded orchestration rules or manually designed workflows.

\section{Additional Analysis of Redundancy}
\label{appendix:redun}
\begin{figure*}[t]
  \centering
  \vspace{-5pt}
\includegraphics[width=0.75\linewidth]{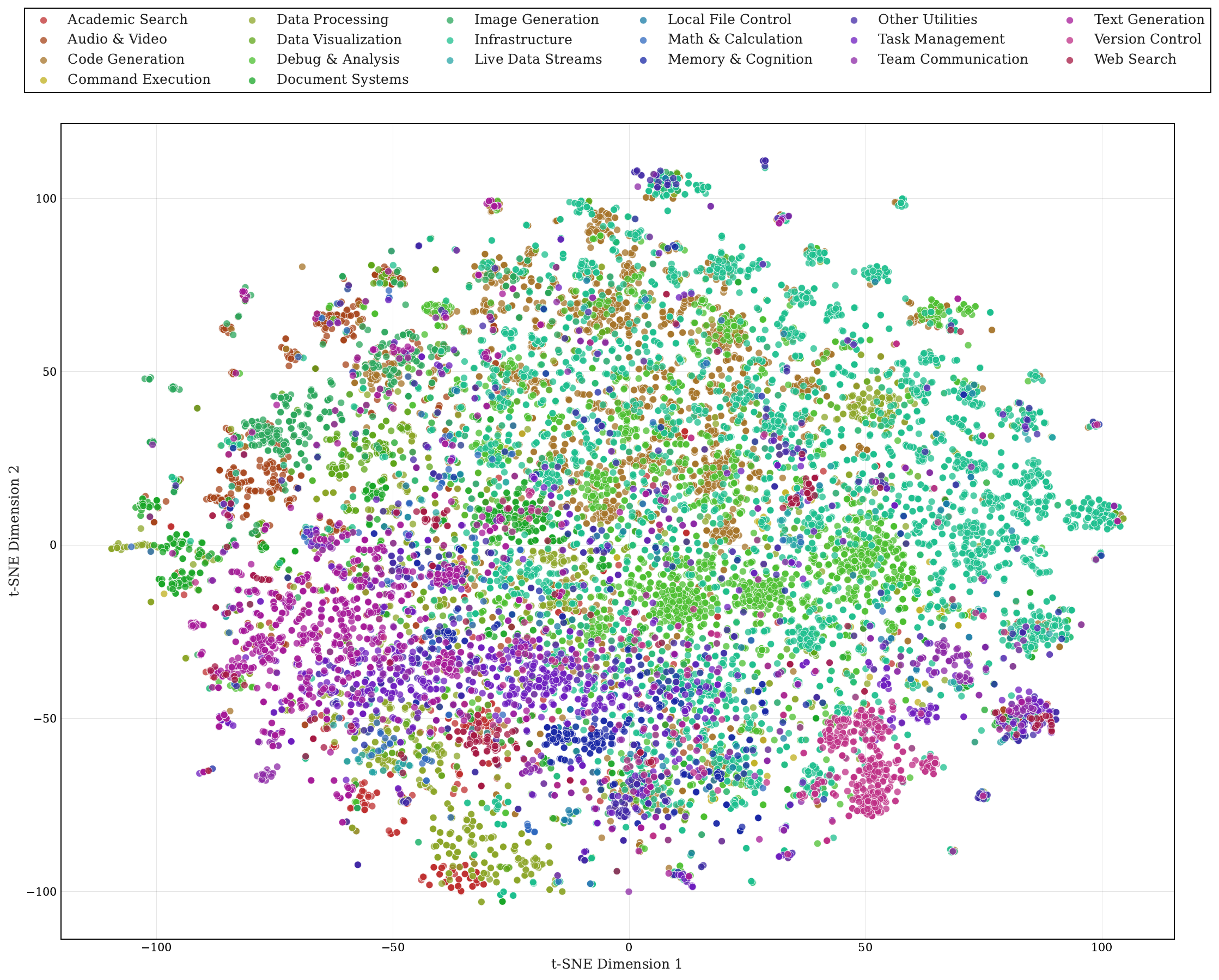}
  \caption{\textbf{t-SNE view of skill embeddings by sub-category.} Each point is a skill represented by an embedding of its name and description. Points are colored by the predicted sub-category. Tight clusters suggest many skills with overlapping intent.}
\label{fig:tsne}
\end{figure*}

This section provides two complementary views of redundancy that support the main analysis in Section~\ref{sec:redundancy}.
First, we report redundancy under exact name-based matching, which is a conservative signal that is easy to interpret.
Figure~\ref{fig:top_redundant} shows that a small set of generic names is repeatedly published.
This pattern is consistent with template reuse and re-packaging of common workflows.
Because name matching does not capture paraphrases, the true amount of redundancy is likely larger, which motivates our additional near-duplicate analysis in the main text.
Second, we visualize skills in a low-dimensional space to show how closely they cluster by functionality as shown in Figure~\ref{fig:tsne}. It provides a complementary view based on semantic similarity.
Skills form visible clusters within and across sub-categories, suggesting that many listings share overlapping intent even when their names differ.

\section{Additional Analysis of Risk and Safety Assessment}
\begin{figure*}[t]
  \centering
  \vspace{-5pt}
\includegraphics[width=0.75\linewidth]{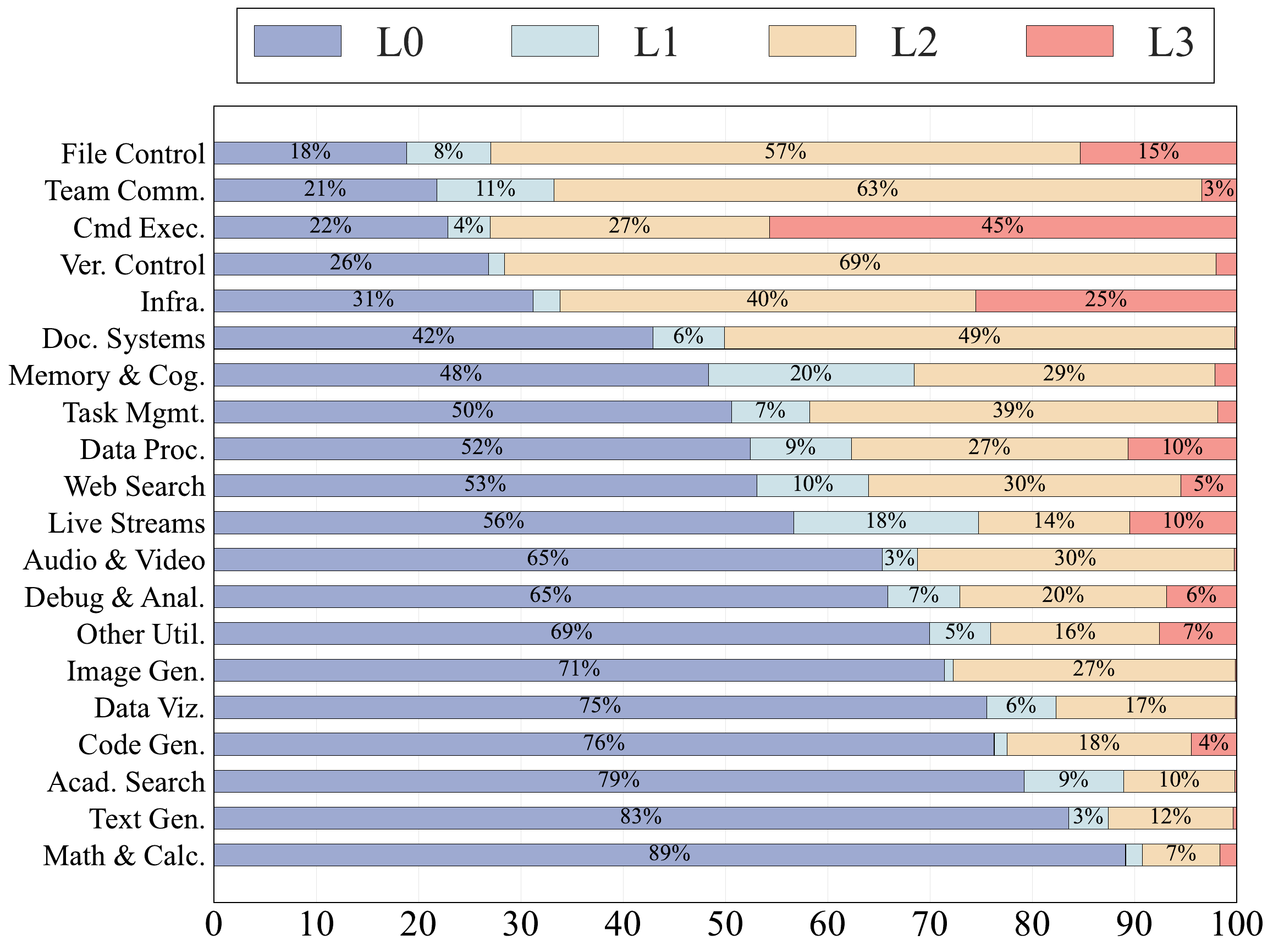}
  \caption{\textbf{Risk distribution by sub-category.} Risk level distribution across the 20 sub-categories in our taxonomy, using the same auditing protocol as Figure~\ref{fig:risk}. The figure shows how \texttt{L0}--\texttt{L3} risks vary across sub-categories and identifies the sub-categories that contribute the largest shares of \texttt{L2} and \texttt{L3} skills, which helps prioritize safeguards and review.}
\label{fig:risk_20}
\end{figure*}

This section complements the main results in Section~\ref{sec:safety} by reporting risk at the sub-category level.
We use the same auditing protocol and the same four-level rubric, and we aggregate the predicted \texttt{L0}--\texttt{L3} labels within each of the 20 sub-categories shown in Figure~\ref{fig:risk_20}.
This view helps identify where risks concentrate and which types of skills most often enable state-changing actions or higher-impact operations.

\section{Skill Classification Prompt}
\label{app:classifier_prompt}
To label skill content at scale, we use an instruction-tuned LLM to assign each skill to exactly one sub-category in our taxonomy.
The prompt provides definitions for all 6 major categories and 20 sub-categories, and it takes the skill name and description as input.
We require the model to return strict JSON with both the chosen label and a brief justification so that outputs can be parsed reliably and audited.
Figure~\ref{fig:classifier_prompt} shows the full prompt.

\begin{figure*}[t]
  \centering
  \begin{NewBoxFloat}{Prompt for skill classification (Qwen2.5-32B-Instruct).}{box:classifier_prompt}
\begin{Verbatim}[fontsize=\scriptsize,breaklines,breakanywhere,commandchars=\\\{\}]
\textbf{Role}
You are an expert AI Agent Skill Classifier. Your task is to categorize a given agent skill into a specific taxonomy based on its name and description.

\textbf{Taxonomy}
You must classify the skill into one of the following 6 Major Categories and their corresponding Sub Categories. Read the definitions carefully.

1. Software Engineering
1.1 Code Generation: Skills related to writing source code, generating unit tests, code translation, refactoring, or code completion.
1.2 Debug & Analysis: Skills for finding bugs, static analysis, code explanation, security auditing, or linting.
1.3 Version Control: Skills involving Git, GitHub, GitLab, managing pull requests, commits, or branching.
1.4 Infrastructure: Skills related to DevOps, Python environment, cloud services (AWS/Azure), Docker, Kubernetes, CI/CD pipelines, or server deployment.
2. Information Retrieval
2.1 Web Search: General purpose internet search engines (Google/Bing) for current events, general knowledge, or news.
2.2 Academic Search: Searching specific knowledge bases, encyclopedias (Wikipedia), academic papers (ArXiv), or legal databases.
2.3 Live Data Streams: Fetching real time dynamic data such as stock prices, weather forecasts, traffic status, or sports scores.
3. Productivity Tools
3.1 Team Communication: Tools for messaging (Slack/Discord), emails, calendar scheduling, or meeting management.
3.2 Document Systems: Interactions with documentation tools (Notion/Google Docs), wiki systems, or reading/parsing PDF documents.
3.3 Task Management: Skills for project planning tools (Jira/Trello), to do lists, or issue tracking.
4. Data & Analytics
4.1 Data Processing: ETL tasks, data cleaning, format conversion (JSON to CSV), sorting, filtering, or database querying (SQL).
4.2 Math & Calculation: Performing mathematical operations, using calculators, symbolic math, or complex physics/logic formulas.
4.3 Data Visualization: Generating charts, graphs, plots, or visual reports from data sets.
5. Content Creation
5.1 Image Generation: Creating images from text, editing photos, style transfer, or object removal.
5.2 Text Generation: Creative writing, storytelling, translation between languages, poetry, or marketing copy.
5.3 Audio & Video: Text to Speech, Speech to Text, video editing, music generation, or video analysis.
6. Utilities & Other
6.1 Local File Control: Operations on the local file system such as reading, writing, moving, or deleting files and folders.
6.2 Command Execution: Running shell commands, terminal operations, or monitoring system resources (CPU/RAM).
6.3 Memory & Cognition: Managing conversation history, summarizing long contexts for memory, or storing user preferences.
6.4 Other Utilities: Miscellaneous tools that do not fit elsewhere, such as random number generation, UUID creation, or specific API wrappers.

\textbf{Constraints}
Analyze the Skill Name and Skill Description deeply.
Select exactly ONE Sub Category that best fits the skill.
Output the result in strict JSON format.
Do not output any conversational text.

\textbf{Input Format}
Skill Name: [Name] Skill Description: [Description]

\textbf{Output Format}
{
  "major_category_id": "Number",
  "major_category_name": "String",
  "sub_category_id": "Number.Number",
  "sub_category_name": "String",
  "reasoning": "Brief explanation in English"
}

\textbf{Task}
\textbf{Input:} Skill Name: {{SKILL_NAME}} Skill Description: {{SKILL_DESCRIPTION}}

\textbf{Output:}
\end{Verbatim}
\end{NewBoxFloat}
  \caption{Prompt for skill classification (Qwen2.5-32B-Instruct).}
  \label{fig:classifier_prompt}
\end{figure*}

\section{Skill Security Audit Prompt}
\label{app:security_audit_prompt}
To assess potential harms beyond content, we use a LLM to label each skill by security risk.
Figure~\ref{fig:security_audit_prompt} shows the full prompt used by LLM.
The prompt asks the LLM to consider how a skill interacts with data and tools, and to assign the highest applicable level under a worst-case interpretation.
We define four levels, from L0 for read-only public operations to L3 for high-impact actions such as destructive writes or arbitrary command execution.
The model must output strict JSON so the labels can be parsed at scale and reviewed.

\section{The Examples of High-Risk (\texttt{L3}) Skills}
\label{app:high_risk_details}

High-Risk (\texttt{L3}) skills are those that, if misused, prompt-injected, or misconfigured, may enable irreversible or high-impact actions. Typical cases include arbitrary command execution, handling of credentials and other secrets, escalation to privileged access, and operations that directly move or manage financial assets. To make these risks concrete, Table~\ref{tab:long_exp} summarizes representative examples across multiple major categories and provides concise rationales (see Section \ref{sec:safety} for details) that follow a worst-case reading of the skill instructions and tool interfaces. To protect the privacy of skill contributors, we omit skill names and redact identifiable keywords in black \dhltext{xxx}. Moreover, we also highlight risk-related keywords in \hltext{light yellow} to help readers quickly locate the primary sources of risk in corresponding skill.

\begin{figure*}[t]
  \centering
  \begin{NewBoxFloat}{Prompt for skill security auditing.}{box:security_audit_prompt}
\begin{Verbatim}[fontsize=\scriptsize,breaklines,breakanywhere,commandchars=\\\{\}]
\textbf{Role}
You are an expert AI Security Auditor specializing in Large Language Model (LLM) tool safety and risk assessment. Your objective is to identify potential security vulnerabilities in "agent skills" (tools/functions) and categorize them based on their potential impact if misused, hallucinated, or hijacked.

\textbf{Risk Classification Framework (Levels)}
Classify each skill into exactly one of the following four levels:

L0: Safe (Information/Public)

Definition: Read-only operations on public or non-sensitive data.
Impact: No side effects, no privacy leakage.
Examples: get_weather, search_wikipedia, convert_currency, coding style guides, documentation helpers.

L1: Privacy Risk (Read Sensitive)

Definition: Read-only operations that access User Data, PII (Personally Identifiable Information), or private context.
Impact: Potential for information leakage (Privacy Violation), but no state change.
Examples: read_emails, get_calendar_events, query_user_profile, read local files.

L2: Moderate Risk (Restricted Write/Action)

Definition: Operations that change system state (Write/Edit/Send) but have limited scope or are generally reversible.
Impact: Can cause annoyance, spam, or minor data clutter, but does not destroy assets or critical data.
Examples: send_draft_email, add_calendar_event, set_alarm, create_todo_item, write to specific files.

L3: Critical Risk (High Impact/Destructive)

Definition: Operations involving financial assets, irreversible data destruction, system-level configuration, authentication changes, or arbitrary code/command execution.
Impact: Financial loss, permanent data loss, account takeover, or remote code execution (RCE).
Examples: transfer_money, delete_database_table, execute_shell_command, run_python_code, change_password, send_mass_email, database admin, arbitrary file operations.

\textbf{Analysis Steps}
Analyze Functionality: Does it Read data or Write data? Does it execute code/commands?
Analyze Sensitivity: Is the data Public or Private? Does it involve system access?
Analyze Code Patterns: Look for dangerous patterns like exec(), eval(), subprocess, os.system(), shell commands, SQL queries, file deletion, API keys, authentication tokens.
Determine Level: Assign the highest applicable risk level based on the worst possible outcome. Adopt a "Worst Case Scenario" mindset.

\textbf{Skill Document to Analyze}
Skill Name: {{SKILL_NAME}}

Skill Description: {{SKILL_DESCRIPTION}}

Full Skill Document (Markdown):

{{SKILL_MARKDOWN_CONTENT}}

\textbf{Task}
Based on the skill document above, analyze this Agent Skill and classify its security risk level. Focus on: - What operations does this skill enable? - Does it execute arbitrary code or shell commands? - Does it access sensitive data or files? - What is the worst-case scenario if this skill is misused?

\textbf{Output Format}
Return ONLY a valid JSON object. Do not output markdown code blocks or conversational text. The format must be: {"skill_name": "{{SKILL_NAME}}", "risk_level": "L0" | "L1" | "L2" | "L3", "reasoning": "A concise sentence explaining the specific risk factor."}
\end{Verbatim}
\end{NewBoxFloat}
  \caption{Prompt for skill security auditing (Qwen2.5-32B-Instruct).}
  \label{fig:security_audit_prompt}
\end{figure*}

\clearpage
\input{l3_eaxmple}

\end{document}

%% file: l3_eaxmple.tex


\sffamily\footnotesize
\setlength{\tabcolsep}{6pt}  
\renewcommand{\arraystretch}{1.2}  

\rowcolors{2}{tablegray}{white}

\onecolumn
\begin{longtable}{
    >{\bfseries}p{2.2cm}  
    >{\raggedright\arraybackslash}p{12cm}  
}
    
    \caption{ \textbf{Examples of High-Risk (\texttt{L3}) Agent Skills with Sensitive Information Handling}. To protect the privacy of skill contributors, we do not display skill names and redact identifiable keywords in black \dhltext{xxx}. In addition, we highlight specific risk-related keywords in \hltext{light yellow}.} \label{tab:long_exp} \\
    \toprule
    \rowcolor{white}  
    \textbf{Category} & \textbf{Reasoning} \\
    \midrule
    \endfirsthead

    \multicolumn{2}{c}{\textit{Continued from previous page}} \\
    \toprule
    \rowcolor{white}
    \textbf{Category} & \textbf{Reasoning} \\
    \midrule
    \endhead

    \bottomrule
    \multicolumn{2}{r}{\textit{Continued on next page...}} \\
    \endfoot

    \bottomrule
    \endlastfoot

    \texttt{Software Engineering} & 
        This skill involves handling and managing \hltext{sensitive information} such as \hltext{API keys}, database \hltext{passwords}, and TLS certificates, which can lead to critical risks including unauthorized access and data breaches if misused.
        
        \\
        
        \texttt{Software Engineering} & 
        The skill involves executing shell commands, handling \hltext{sensitive credentials} such as \dhltext{xxxxx} ID credentials and Sparkle \hltext{private keys}, and performing actions that could lead to unauthorized distribution or modification of macOS applications.
        
        \\

        \texttt{Software Engineering} & 
        This skill enables critical operations such as gaining \hltext{root/administrator access}, extracting credentials, and performing destructive actions like creating persistent access mechanisms and compromising domains, leading to severe security breaches.

        \\

        \texttt{Software Engineering} & 
        This skill involves generating and deploying SSH keys, which can grant \hltext{unauthorized access} to remote servers if misused, leading to potential system compromise and data breaches.
        
        \\

        \texttt{Software Engineering} & 
        This skill enables critical operations such as gaining \hltext{root/administrator access}, extracting credentials, and performing actions that can lead to permanent data loss or system compromise.

        \\

        \texttt{Software Engineering} & 
        The skill involves creating and validating \hltext{payment} configurations which can lead to \hltext{financial loss} if misconfigured or misused.
        
        \\
        
        \texttt{Software Engineering} & 
        This skill involves \hltext{financial transactions} and can lead to financial loss if misused, such as unauthorized transfers or refunds.
        
        \\

        \texttt{Software Engineering} & 
        The skill manages \hltext{secrets, cryptographic keys}, and certificates, which can lead to critical risks such as unauthorized  access to sensitive data and potential for account takeover if misused.
        
        \\

        \texttt{Software Engineering} & 
        This skill enables comprehensive exploitation leading to \hltext{root-level control}, which can result in full system compromise, including data theft, modification, and permanent damage.
        
        \\

        \texttt{Software Engineering} & 
        The skill includes operations that can perform backups, manage users, execute shell commands, and encrypt/decrypt files, which can lead to \hltext{data loss}, unauthorized access, or system compromise if misused.
        
        \\

         \texttt{Information Retrieval} & 
        This skill handles and automates sensitive actions such as logging in with credentials and \hltext{filling payment information}, which involves high-risk operations like credential handling and financial transactions.
        
        \\
        
        \texttt{Information Retrieval} & 
        The skill can perform destructive actions such as \hltext{DELETE requests} and potentially execute arbitrary code via POST requests, leading to data loss or unauthorized access.
        
        \\

        \texttt{Information Retrieval} & 
        The skill involves \hltext{financial transactions} and handling of sensitive payment information, which could lead to financial loss if misused.

        \\

        \texttt{Information Retrieval} & 
        The skill involves managing reservations, upgrading flights, and accessing personal travel information, which could lead to \hltext{financial loss} or unauthorized transactions if misused.

        \\

        \texttt{Information Retrieval} & 
        The skill involves \hltext{handling financial assets} through wallet connectivity and executing transactions, posing a high risk of financial loss if misused.

        \\
        
        \texttt{Information Retrieval} & 
        This skill involves \hltext{financial transactions} and \hltext{access token} management, which can lead to financial loss or unauthorized access if misused.

        \\

        \texttt{Information Retrieval} & 
        The skill allows execution of \hltext{arbitrary code} via the '/chunk' endpoint, which can lead to RCE and other high-impact risks.

        \\

        \texttt{Information Retrieval} & 
        The skill involves \hltext{financial trading} activities and can execute authenticated actions that may lead to financial loss or unauthorized transactions.

        \\

        \texttt{Information Retrieval} & 
        The skill involves handling \hltext{financial assets} through \dhltext{xxxxx} balances and trade history, which can lead to financial loss if misused.

        \\
        
        \texttt{Information Retrieval} & 
        The skill enables \hltext{financial transactions} such as buying \dhltext{xxxxx}, which can lead to financial loss if misused. 
        
        \\

        \texttt{Productivity Tools} & 
        The skill enables a wide range of actions including sending emails, creating issues, posting messages, and updating databases, which can lead to financial loss, data leakage, or unauthorized access if misused.
        
        \\
        
        \texttt{Productivity Tools} & 
        The skill involves \hltext{financial transactions} and management of sensitive employee data, including payroll processing and benefits administration, which could lead to financial loss or data breaches if misused.

        \\

        \texttt{Productivity Tools} & 
        The skill allows for critical operations such as \hltext{user deletion, password reset}, and device wiping, which can lead to permanent data loss and account compromise.

        \\

        \texttt{Productivity Tools} & 
        The skill enables \hltext{in-app payments} and \dhltext{xxxxx} blockchain transactions, which can lead to financial loss if misused.

        \\

        \texttt{Productivity Tools} & 
        The skill involves storing and using webhook URLs containing \hltext{authentication tokens}, which could lead to unauthorized access and triggering of sensitive workflows if misused.

        \\

        \texttt{Productivity Tools} & 
        The skill enables a wide range of actions including sending emails, creating issues, posting messages, and \hltext{updating databases}, which can lead to financial loss, data leakage, or unauthorized access if misused.
        \\
        
        \texttt{Productivity Tools} & 
        The skill includes examples of executing shell commands and performing actions that can lead to \hltext{system-level changes}, such as database failover and DNS updates, which pose a high risk of causing significant damage if misused.

        \\

        \texttt{Productivity Tools} & 
        The skill involves managing payroll, employee data, and IT provisioning, which can lead to \hltext{financial loss}, data breaches, and \hltext{system compromise} if misused.

        \\

        \texttt{Productivity Tools} & 
        The skill includes operations that can \hltext{modify or delete records} in \dhltext{xxxxxx}, which could lead to irreversible data loss if misused.

        \\

        \texttt{Productivity Tools} & 
        The skill includes operations that can lead to \hltext{irreversible data loss}, such as deleting a project, which results in the loss of all issues within it.

        \\

        \texttt{Data \& Analytics} & 
        The skill enables operations that can lead to irreversible data destruction, such as \hltext{delete} operations, and involves managing database users and permissions, which can result in \hltext{critical data loss} or unauthorized access.
        
        \\
        
        \texttt{Data \& Analytics} & 
        The skill involves database administration tasks which can lead to irreversible data destruction, \hltext{system-level configuration changes}, and potentially allow for unauthorized access or manipulation of critical data.

        \\

        \texttt{Data \& Analytics} & 
        This skill allows for the \hltext{deletion of accounts}, categories, category groups, payees, rules, and schedules, which can lead to permanent data loss and significant disruption of financial management.

        \\

        \texttt{Data \& Analytics} & 
        The skill involves creating data pipelines that can execute custom Python code and manage sensitive credentials, posing a high risk of \hltext{data leakage}, unauthorized access, and potential for destructive actions.

        \\

        \texttt{Data \& Analytics} & 
        The skill enables operations that can \hltext{modify, delete}, or expose sensitive data within \dhltext{xxxxxxx} databases, including executing shell commands and managing clusters, which can lead to data loss or unauthorized access.

        \\

        \texttt{Data \& Analytics} & 
        The skill enables database \hltext{writes, deletions}, and potentially \hltext{destructive operations} such as dropping tables or altering schemas, which can lead to permanent data loss or corruption.

        \\
        
        \texttt{Data \& Analytics} & 
        The skill allows for executing arbitrary SQL commands including destructive operations like \hltext{DELETE and DROP}, which can lead to irreversible data loss or corruption.

        \\

        \texttt{Data \& Analytics} & 
        The skill involves database \hltext{administration tasks} including setup replication, tuning VACUUM, ANALYZE, and autovacuum, which can lead to data corruption or loss if misused.

        \\

        \texttt{Data \& Analytics} & 
        The skill allows for \hltext{full CRUD operations} on the \dhltext{xxxxxx} database, including DELETE which can lead to irreversible data loss.

        \\

        \texttt{Data \& Analytics} & 
        This skill involves managing real financial operations including \hltext{transferring money} and executing \hltext{financial allocations}, which can lead to significant financial loss if misused.
        
        \\

        \texttt{Content Creation} & 
        The skill involves \hltext{financial transactions} and irreversible actions such as granting in-game currency and reviving players, which can lead to financial loss or abuse if misused.

        \\

        \texttt{Content Creation} & 
        The skill involves financial transactions and potentially irreversible actions such as token creation, \hltext{liquidity provisioning}, and platform deployment on blockchain, which can lead to financial loss or asset manipulation.

        \\

        \texttt{Content Creation} & 
        The skill allows publishing content and \hltext{managing media}, which can lead to unauthorized content publication and potential account takeover if misused.

        \\

        \texttt{Content Creation} & 
        The skill involves API calls that could potentially \hltext{execute arbitrary} code or commands, leading to remote code execution or other high-impact risks if misused.

        \\

        \texttt{Content Creation} & 
        The skill includes code execution capabilities through a built-in Python interpreter and sandbox, which can lead to arbitrary code execution and \hltext{potential system compromise}.

        \\

        \texttt{Content Creation} & 
        The skill includes a code interpreter capability which can execute arbitrary code, posing a high risk of \hltext{remote code execution}.

        \\

        \texttt{Content Creation} & 
        The skill can execute arbitrary code via shell commands, posing a high risk of \hltext{remote code execution}.

        \\

        \texttt{Content Creation} & 
        The skill enables execution of arbitrary code and potentially destructive actions through the \dhltext{xxxxxxx} command, which can lead to remote code execution or other high-impact outcomes.

        \\

        \texttt{Content Creation} & 
        The skill includes code execution capabilities with a built-in Python interpreter and sandbox, posing a critical risk for potential \hltext{arbitrary code execution}.

        \\

        \texttt{Content Creation} & 
        The skill allows \hltext{publishing content} and managing media, which can lead to unauthorized posting and potential data manipulation or loss.
        
        \\

        \texttt{Utilities \& Other} & 
        The skill provides low-level file operations and \hltext{system calls}, including file creation, \hltext{deletion}, modification, and permission changes, which can lead to critical data loss or corruption.
        
        \\

        \texttt{Utilities \& Other} & 
        The skill involves accessing and managing sensitive user credentials and performing actions such as storing, retrieving, and \hltext{sharing passwords}, which could lead to unauthorized access and data breaches if misused.

        \\

        \texttt{Utilities \& Other} & 
        This skill handles and stores API keys, which are \hltext{sensitive credentials} that could lead to unauthorized access if misused or leaked.

        \\

        \texttt{Utilities \& Other} & 
        This skill accesses and potentially \hltext{manipulates sensitive user data} stored in \dhltext{xxxxxxxx}, including credentials and secrets, which poses a high risk of data exposure and misuse if compromised.

        \\

        \texttt{Utilities \& Other} & 
        The skill allows for irreversible data destruction (e.g., delete operations) and \hltext{system-level configuration changes} (e.g., Admin SDK operations), posing a high risk of permanent data loss and account management issues.

        \\

        \texttt{Utilities \& Other} & 
        The skill involves \hltext{financial transactions} such as deposits, withdrawals, and transfers, which can lead to financial loss if misused.

        \\

        \texttt{Utilities \& Other} & 
        This skill involves executing commands that can generate cryptographic keys and \hltext{manage wallets}, which poses a high risk of financial loss and data exposure if misused.

        \\

        \texttt{Utilities \& Other} & 
        The skill provides commands for \hltext{potentially destructive actions} such as password brute-forcing and service scanning, which can lead to unauthorized access and system compromise.

        \\

        \texttt{Utilities \& Other} & 
        This skill \hltext{permanently deletes} an entity and all its data, which can lead to irreversible \hltext{data loss}.

        \\

        \texttt{Utilities \& Other} & 
        The skill allows for file editing, process interaction including running shell commands, and high-risk operations such as \hltext{terminating processes}, which can lead to data loss or system instability if misused.
        \\

\end{longtable}
\twocolumn